\documentclass[12pt,epsf]{article}
\usepackage[pctex32]{graphics}
\makeatletter

\@addtoreset{equation}{section}
\newcommand{\be}{\begin{eqnarray}}
\newcommand{\ee}{\end{eqnarray}}
\newcommand{\ba}{\begin{array}}
\newcommand{\ea}{\end{array}}
\newcommand{\nn}{\nonumber}

\makeatletter \@addtoreset{equation}{section} \makeatother

\begin{document}
\vspace{1cm}
\begin{center}
~\\~\\~\\
{\bf  \LARGE Generalized Gravitational Entropy of  Interacting Scalar Field and Maxwell Field }
\vspace{1cm}

                      Wung-Hong Huang\\
                       Department of Physics\\
                       National Cheng Kung University\\
                       Tainan, Taiwan\\
\end{center}
\vspace{1cm}
\begin{center}{\bf  \Large ABSTRACT } \end{center}
The generalized gravitational entropy proposed by Lewkowycz  and Maldacena in recent is extended to the interacting real scalar field and Maxwell field system.  Using the BTZ geometry we first investigate the case of free real scalar field and then show a possible way to calculate the entropy of the interacting scalar field. Next, we investigate the Maxwell field system.  We exactly solve the wave equation and calculate the analytic value of  the generalized gravitational entropy.  We also use the Einstein equation to find the effect of backreaction of the Maxwell field on the area of horizon.  The associated modified area law is consistent with the generalized gravitational entropy.  

\vspace{3cm}
\begin{flushleft}
*E-mail:  whhwung@mail.ncku.edu.tw\\
\end{flushleft}
\newpage
\section{Introduction}
In recent  Lewkowycz  and Maldacena [1] proposed the generalization of the usual black hole entropy formula [2-5] to Euclidean solutions without a Killing vector. Let us briefly describe it in below.
 
For a general quantum system the Euclidean evolution generates an un-normalized density matrix
\be
\rho = P~e^{-\int^{\tau_f}_{\tau_0} d\tau H(\tau)}
\ee
While $\rm Tr[\rho]$ can be computed by considering Euclidean evolution on a circle by identifying $\tau_f=\tau_0+2\pi$.  The entropy of this density matrix can be calculated by the ``replica trick" [6,7].   Namely, we can compute $\rm Tr[\rho^n]$ by considering time evolution over a circle of n times the length of the original one  where  $H(\tau+2\pi) = H(\tau)$ and entropy is computed as
\be
S&=&-n\partial_n\Big[\log Z(n)-n \log Z(1)\Big]_{n=1}=\rm Tr\Big[\hat\rho\log\hat\rho\Big]\\
Z(n)&=&\rm Tr\Big[\rho^n\Big],~~~~~\hat\rho={\rho\over \rm Tr[\rho]}
\ee
In the gravitational context, we can consider metrics which end on a boundary and assume that the boundary has a direction with the topology of the circle.
The boundary data can depend on the position along this circle but it respects the periodicity of the circle. We define the coordinate $\tau\sim \tau+2\pi$ on the circle but without a U(1) symmetry. We then consider a spacetime in the interior which is smooth. Its Euclidean action is defined to be log Z(1). We can also consider other spacetimes where we take the same boundary data but consider a new circle with period $\tau\sim \tau+2n\pi$. Their Euclidean action is defined to be log Z(n). These computations can be viewed as computing$\rm Tr[\rho^n]$ for the density matrix produced by the Euclidean evolution. 

In this way,  Lewkowycz  and Maldacena [1] propose the generalization of the usual black hole entropy formula [2-5] to Euclidean solutions without a Killing vector.  While the original paper had used the complex scalar field to calculate the black hole entropy we will first investigate the case of free real scalar field and then discuss the effect of the interacting scalar field and the Maxwell field. 

In section II we study the free real scalar field and interacting scalar field.  In section III we exactly solve the Maxwell field equation and calculate the analytic value of  the generalized gravitational entropy.  In section IV we use the Einstein equation to study the effect of backreaction of the Maxwell field on the spacetime.  We calculate the associated modified area law and see that it is consistent with the generalized gravitational entropy. Last section is devoted to a short conclusion.
\section {Generalized Gravitational Entropy of Interacting Real Scalar Field}
\subsection {Geometry} 
We follow the original paper [1] and consider the simple BTZ geometry \footnote{This geometry is more like $\rm AdS^3$ then the BTZ spacetime.} 
\be
ds^2={dr^2\over 1+r^2}+r^2d\tau^2+(1+r^2)dx^2
\ee
Above metric has a U(1) isometry along the circle labeled by $\tau\sim \tau+2\pi$. As metric is invariant under translations in $x$ this direction will not play any role and we can take it to be compact of size $L_x$. Computing the entropy for this solution gives the standard area formula.

We now add a scalar field or Maxwell field and set boundary conditions that are not U(1) invariant by
\be
\Phi&\sim& \eta\cos(\tau)~~~~~~~~~~~~\rm~~at~~~r\rightarrow\infty\\
A_\mu&\sim& \eta\cos(\tau)~~~~~~~~~~~~\rm~~at~~~r\rightarrow\infty
\ee
For the $n^{th}$ case, we need to consider a spacetime with the same boundary
conditions in above equation but where $\tau\sim \tau+2n\pi$. This implies that the spacetime in the interior is [1] 
\be
ds^2={dr^2\over n^{-2}+r^2}+r^2d\tau^2+(n^{-2}+r^2)dx^2
\ee
We now need to compute the gravitational action to second order in $\eta$ for scalar field or Maxwell field.  The metric is changed at order $\eta^2$ but since the original  background obeys Einstein equations, there is no contribution from the gravitational term to order $\eta^2$. Thus, to this order, the whole contribution comes from the scalar field  or  Maxwell field in the action and we need to consider it in this spacetime. 
\subsection {Free Real Scalar Field}
The action of real scalar field is
\be
A^{\Phi}={1\over 2}\int d^3x\sqrt g~g^{ab}\partial_a\Phi \partial_b\Phi={1\over 2}\int d^3x\partial_a[\sqrt g~g^{ab}\Phi \partial_b\Phi]-{1\over 2}\int d^3x \Phi\partial_a[\sqrt g~g^{ab} \partial_b\Phi]
\ee
The first bracket is the surface term and will contribute to the gravitational action which is considered later.  After the variation the second bracket gives the scalar field equation. As in [1] we make following ansatz 
\be
\Phi(\tau,r)&=&\eta  \cos(\tau)~\phi_n(r)
\ee
then the associated differential equation of $\phi_n(r) $ becomes 
\be
(n^{-2}r^2+r^4)\phi''_n(r)+(n^{-2}r+3r^3)\phi'_n(r)-\phi_n(r)=0
\ee 
which is just the differential equation of the complex scalar field investigated in the original paper [1].  There are two independent solutions which have been normalized by $\phi_n(\infty)=1$ are 
\be
h_n(r)&=&{n^nr^n \Gamma\Big(1+{n\over 2}\Big)^2\over \Gamma(1+n)}~_2F_1\Big(1+{n\over 2},{n\over 2},1+n,-n^2r^2\Big)\\
k_n(r)&=&{n^{-n}r^{-n} \Gamma\Big(1-{n\over 2}\Big)^2 \over \Gamma(1-n)}~_2F_1\Big(1-{n\over 2},-{n\over 2},1-n,-n^2r^2\Big)
\ee 
The first solution which  is regular at the origin will be used in this section to find the associated gravitational action. The second solution which is divergent at the origin plays a role in next subsection in studying the interacting theory.

 Note that the action, as we analyzed before, has two parts.  The second bracket becomes zero after putting the field on-shell and we remain only the first bracket which is the surface term. Thus the classical on-shell  action becomes 
\be
A_{\rm on-shell}^{\Phi}&=&{1\over 2}\int d^3x\partial_a[\sqrt g~g^{ab}\Phi \partial_b\Phi]={1\over 2}\int d^3x  \Big[\partial_\tau(\sqrt g~g^{\tau\tau}\Phi \partial_\tau\Phi)+\partial_r(\sqrt g~g^{rr}\Phi \partial_r\Phi)\Big]\nn\\
&=&-{\eta^2\over 2}\int d^3x\Big[\partial_t(\cos(\tau)\sin(\tau))\Big(\sqrt g~g^{\tau\tau}h_n(r)^2\Big)\nn\\
&&~~~+\cos^2(\tau)\partial_r\Big(h_n(r)\sqrt g~g^{rr}\partial_rh_n(r)\Big)\Big]\nn\\
&=&{\eta^2\over 2} \int_0^{L_x} dx~\int_0^{2n\pi} d\tau \cos^2(\tau)~\Big(h_n(r)\sqrt g~g^{rr}\partial_rh_n(r)\Big)_{r\rightarrow\infty}\nn\\
&=&{n\pi L_x\over 2}~\eta^2 \Big(h_n(r)\sqrt g~g^{rr}\partial_rh_n(r)\Big)_{r\rightarrow\infty}
\ee
Substituting the solution $h_n(r)$ we find that
\be
\log Z^{\Phi}(n)=A_{\rm on-shell}^{\Phi}&=&\pi  L_x~\eta^2\left(-2 n (\log (n)+\log (r))+2 n \left(\psi    ^{(0)}\left(\frac{n}{2}\right)+\gamma \right)+2\right)
+ {\cal O} \Big({1\over r}\Big)\nn\\
\ee
The terms linear in $n$ include divergent terms that should be subtracted [1]. However, they do not contribute to the entropy. 
 
The associated generalized gravitational entropy becomes 
\be
S^{\Phi}_{GGE}&=&-n\partial_n\Big[\log Z(n)-n \log Z(1)\Big]_{n=1}={\pi  L_x \over 4}~\eta^2\Big(8-\pi^2\Big)
\ee
which is exactly the half value of the complex scalar field calculated in [1], as expected.
\subsection {Interacting Real Scalar Field}
The action of interacting real scalar field is
\be
A&=&{1\over 2}\int d^3x \sqrt g~\Big[g^{ab}\partial_a\Phi \partial_b\Phi+{\lambda\over 4!}\Phi^4\Big]\nn\\
&=&{1\over 2}\int d^3x \partial_a[\sqrt g~g^{ab}\Phi \partial_b\Phi]-{1\over 2}\int d^3x~ \Big[\Phi\partial_a(\sqrt g~g^{ab} \partial_b\Phi)-{\lambda\over 4!}\sqrt g~\Phi^4\Big]
\ee
The first integration is the surface term and will contribute to the gravitational action. After substituting previous ansatz $\Phi(\tau,r)=\eta  \cos(\tau)~F_n(r)$ into second integration it becomes 
\be
(n\pi L_x\eta^2)\Big[\int dr~{1\over 2}F_n(r) \sqrt g~g^{\tau\tau}F_n(r)- F_n(r)\partial_r[\sqrt g~g^{rr}\partial_rF_n(r)]\nn\\
+\int dr~{3\over 4}~{\lambda\eta^2\over 4!}\sqrt g~F_n(r)^4\Big]
\ee
in which ${3\over 4}$ is coming from the integration of $\cos(\tau)^4$ in ${\lambda\over 4!}\Phi^4$.  The variation with respective to $F_n(r)$ gives the differential equation
\be
(n^{-2}r^2+r^4)F''_n(r)+(n^{-2}r+3r^3)F'_n(r)-F_n(r)={\lambda\eta^2\over 8}F^3_n(r)
\ee
which cannot be solved exactly.  Therefore, we restore to the perturbation for small coupling $\lambda\eta^2$.  In this case we expand $F_n=h_n+\lambda \eta^2s_n$ in which $h_n$ is that found in free case and  $s_n$ will satisfy the differential equation
\be
s''_n(r)+{n^{-2}r+3r^3\over n^{-2}r^2+r^4}~s'_n(r)-{1\over n^{-2}r^2+r^4}~s_n(r)={h^3_n(r)\over 8(n^{-2}r^2+r^4)}
\ee
Above differential equation can be solved by the standard method of variation of parameters. As associated homogeneous differential equation has two exact solutions  ($k_n(r)$ and $h_n(r)$) the general solution is
\be
s_n(r)=\alpha~k_n(r)+\beta~h_n(r) + P_n(r) k_n(r)+Q_n(r) h_n(r)
\ee
in which $\alpha$ and $\beta$ are arbitrary constants and 
\be
P_n&=&\int dr{- h_n(r) \over W(k_n(r),h_n(r))}~{h^3_n(r)\over 8(n^{-2}r^2+r^4)}
\\
Q_n&=&\int dr{ k_n(r) \over W(k_n(r),h_n(r))}~{h^3_n(r)\over 8(n^{-2}r^2+r^4)}
\ee
where $W(k_n(r),h_n(r))$ is the Wronskian of $k_n(r), h_n(r)$ defined by
\be
W(k_n(r),h_n(r))&=&\left| {\begin{array}{cc}
k_n(r)& h_n(r) \\
k_n'(r)& h_n'(r)\\
  \end{array} } \right|
\ee
While we could not find the exact function forms of $P_n$ and $Q_n$ we make following comments :

1. In the convention,  we can ignore the  integration constants in above integrations since including them would merely regenerate terms already present in the homogeneous  solution.  However, in our case we shall take $\alpha=0$ and $\beta=1$ to have regular solution at $r=0$.  

2.  Although solution $k_n(r)$ is divergent at $r=0$ the particular solution $P_n$ and $Q_n$, which involve $k_n(r)$ in their definition, are finite at $r=0$.  We have checked this property  by numerical evaluation. Thus we shall take both solutions in following study.

3.  We have to find the large $r$ behaviors of particular solution $P_n(x)$ and $Q_n(x)$ in order to calculate the generalized gravitational entropy. By Taylor expansion about $r=\infty$ it is easy to find that
\be
P_n(r)&\approx&{1\over 12 \pi  n}\tan \Big(\frac{\pi  n}{2}\Big) \Big(-2 \log (r) \Big(n\Big(\log \Big(n^2\Big)+\log (r)\Big)-2 \gamma n+n-1\Big)+n r^2\nn\\
&&+n \Big(\psi ^{(0)}\Big(-\frac{n}{2}\Big)+3 \psi ^{(0)}\Big(\frac{n}{2}\Big)\Big) \log (r)\Big)\\
Q_n(r)&\approx&{1\over 12 \pi  n} \tan \Big(\frac{\pi  n}{2}\Big) \Big(-2 \log (r) \Big(n\Big(\log \Big(n^2\Big)+\log (r)\Big)+n-2\Big)+n r^2\nn\\
&&+4 n \Big(\psi ^{(0)}\Big(\frac{n}{2}\Big)+\gamma \Big) \log (r)\Big)
\ee
Because $P_n$ and $Q_n$ are divergent at $r\rightarrow\infty$ we have to introduce a cutoff $\Lambda$ and perform the integration from $\Lambda$ to large $r$ to define the precisely values of $P_n$ and $Q_n$ in large $r$. In this way the solution is normalized to be $\Phi (\infty)=1$ as in free case.

Now, we can substitute above solutions into classical on-shell  action and the associated generalized gravitational entropy becomes 
\be
S^{\lambda\Phi}_{GGE}&=&-n\partial_n\Big[\log Z(n)-n \log(1)\Big]_{n=1}\approx \pi  L_x ~\Big(8-\pi^2\Big) \Big({\eta^2\over 4}-{\lambda\eta^4\over 64}\Big)\ee
to the first order of small parameter $\lambda\eta^4$.

Note that above result seems to be valid at order $\eta^4$ which  is a higher order than initially expected and  to this order the purely gravitational part of the action  should give a contribution.  Now we can consider the case with the property : next leading gravitational part $<$ interacting scalar field part $<$ leading gravitational part,  i.e. $\eta^4<\lambda \eta^4<\eta^2$ then above result is consistent without furthermore gravitational part.  We shall notice that in this case, as $\eta^4<\lambda\eta^4$ the coupling $\lambda$ is larger than 1 while $\eta < 1$.
\subsection {Backreaction of Interacting Scalar Field on the Area Law}
Now we will study the backreaction of the interacting scalar field on the metric.   Note that the backreaction of free scalar field has been treated in original paper [1].  Following it we can easily check that the formula derive in [1]
\be
S^{\Phi}|_{\eta^2}&=&4\pi (\delta A_0+\delta A_{\lambda\eta^2})=4\pi \delta A_0+S^{\Phi}|_{\lambda\eta^2}= -\lambda\eta^2~2\pi L_x  \int_0^\infty dr  r(\partial_r\Phi)^2
\ee
can also be used in the interacting scalar system.  The black hole entropy is $S = 4\pi A = 4\pi (A_0 +\delta A_0+\delta A_{\lambda\eta^2})$ where $ 4\pi \delta A_0$ is the part of deformation by  free scalar field which was studied in [1].  $4\pi \delta A_{\lambda\eta^2}$ is the part of deformation by  interacting scalar field.  

The field $\Phi$ in above equation can be found by the standard method of variation of parameters as described in previous subsection.  Now we need to find the $n=1$ solutions associated to $h_n(r)$ and $k_n(r)$ in (2.8) and (2.9).  To avoid the singularity of factor $~_2F_1\left(1-\frac{n}{2},-\frac{n}{2};1-n;-n^2 r^2\right)$ in $k_n(r)$ in the limit $n=1$ we can use following identity 
\be
(1-n)~_2F_1\left(1-\frac{n}{2},-\frac{n}{2};1-n;-n^2 r^2\right) = \left(1-\frac{n}{2}\right)^2 \, _2F_1\left(2-\frac{n}{2},-\frac{n}{2};2-n;-n^2
   r^2\right)\nn\\
-\frac{n^2}{4}  \, _2F_1\left(1-\frac{n}{2},1-\frac{n}{2};2-n;-n^2 r^2\right)~~~~~~~~\ee
to find  $k_1(r)$.  Then we see that  $h_1(r)=k_1(r)$ although  $h_n(r)\ne k_n(r)$ if $n\ne 1$.  In fact for case of $n=1$ we can find the following exact solutions 
\be
h_1(r)&=&\frac{\pi  r}{4}  \, _2F_1\left(\frac{1}{2},\frac{3}{2};2;-r^2\right)\\
\ell_1(r)&=&-\frac{i ~{\bf E}\left(r^2+1\right)}{r}=h_1(r) +~ i~F(r)
\ee
where ${\bf E(r)}$ is the complete elliptic integrals of the second kind.  Note that functions  $h_1(r)$ and $F(r)$ are real. $h_1(\infty)=\ell_1(\infty)=1$ and  $F(\infty)=0$.  Therefore, besides the function  $\ell_1(r)$ we can choose $h_1(r) + \alpha~F(r)$ as another real solution and both of them satisfy the boundary condition of approach to 1 at $r=\infty$.  However, we now have an arbitrary parameter $\alpha$. Maybe, for example, we can set the boundary condition at  $r=10$ by  requiring the solution to approach 0.1 to determine the parameter $\alpha$. However this will render the analytic calculation in previous subsection to have another arbitrary parameter. 

 Therefore at present time we can not see the property that the backreaction of the interacting scalar  field on the area of horizon is equal to the associated generalized gravitational entropy, which is show in the case of free scalar field in [1].  In next section we can see that the backreaction of the Maxwell  field on the area of horizon is exactly equal to the associated generalized gravitational entropy. 
\section {Generalized Gravitational Entropy  of Maxwell Field}
The conventional action of Maxwell field is
\be
A=-{1\over 4}\int d^3x\sqrt g~F_{ab}F_{cd}g^{ac}g^{bd}
\ee
We choose the gauge of $A_0=0$ and as [1] we assume that the Maxwell fields  are independent of coordinate $x$.  Then, the action becomes
\be
A&=&-{1\over 2}\int d^3x\sqrt g~\Big(g^{\tau\tau}g^{xx}\Big(\partial_tA_x\Big)^2+g^{rr}g^{xx}\Big(\partial_rA_x\Big)^2+g^{\tau\tau}g^{rr}\Big(\partial_tA_r\Big)^2\Big)\\
&=&-{1\over 2}\int d^3x \Big[\partial_t\Big(A_x\sqrt g~g^{\tau\tau}g^{xx}\partial_tA_x\Big)+\partial_r\Big(A_x\sqrt g~g^{rr}g^{xx}\partial_rA_x\Big)\nn\\
&& + \partial_t\Big(A_r\sqrt g~g^{\tau\tau}g^{rr}\partial_tA_r\Big)\Big] -\Big[A_x \partial_t\Big(\sqrt g~g^{\tau\tau}g^{xx}\partial_tA_x\Big)\nn\\
&&+A_x \partial_r\Big(\sqrt g~g^{rr}g^{xx}\partial_rA_x\Big)+ A_r \partial_t\Big(\sqrt g~g^{\tau\tau}g^{rr}\partial_tA_r\Big)\Big]
\ee
The first bracket is the surface term and will contribute to the gravitational action which is considered later.  After the variation the second bracket gives the field equations (as in [1] we assume that Maxwell field is independent of $x$)
\be
0&=&\partial_t\Big(\sqrt g~g^{\tau\tau}g^{rr}\partial_tA_r\Big)
\\
0&=&\partial_t\Big(\sqrt g~g^{\tau\tau}g^{xx}\partial_tA_x\Big)+\partial_r\Big(\sqrt g~g^{rr}g^{xx}\partial_rA_x\Big)
\ee
If we search the following solution 
\be
A_r(\tau,r)&=&\eta_r \cos(\tau)~w_n(r)\\
A_x(\tau,r)&=&\eta~ \cos(\tau)~f_n(r)
\ee
then field equations imply $w_n(r)=0$ and $f_n(r)$ is described by 
\be
rf''_n(r)+f'_n(r)-{1\over r(n^{-2}+r^2)}f_n(r)=0
\ee 
The solution which  is regular at the origin is
\be
f_n(r)=n^nr^n~_2F_1\Big({n\over 2},{n\over 2},{n\over 2},1+n,-n^2r^2\Big)
\ee
As in real scalar field we choose the normalized constant to be 1 for the coefficient of leading asymptotical term.  Then the normalized solution  becomes
\be
f_n(r)={n^nr^n \Gamma\Big({n\over 2}\Big)\Gamma\Big(1+{n\over 2}\Big)\over 2 n\Gamma(n)}~_2F_1\Big({n\over 2},{n\over 2},{n\over 2},1+n,-n^2r^2\Big)
\ee 
Using this solution we now calculate the gravitational action. 

 The action we analyzed before has two parts.  The second bracket becomes zero after putting the field on-shell and we remain only the first bracket which is the surface term. Thus the classical on-shell  action becomes 
\be
A_{\rm on-shell}&=&-{1\over 2}\int d^3x \Big[\partial_t\Big(A_x\sqrt g~g^{\tau\tau}g^{xx}\partial_tA_x\Big)+\partial_r\Big(A_x\sqrt g~g^{rr}g^{xx}\partial_rA_x\Big)\Big]\nn\\
&=&-{\eta^2\over 2}\int d^3x\Big[\partial_t(-\cos(\tau)\sin(\tau))\Big(\sqrt g~g^{\tau\tau}g^{xx}f_n(r)^2\Big)\Big]\nn\\
&&~~~+\cos^2(\tau)\partial_r\Big(f_n(r)\sqrt g~g^{rr}g^{xx}\partial_rf_n(r)\Big)\nn\\
&=&-{\eta^2\over 2} \int_0^{L_x} dx~\int_0^{2n\pi} d\tau \cos^2(\tau)~\Big(f_n(r)\sqrt g~g^{rr}g^{xx}\partial_rf_n(r)\Big)_{r\rightarrow\infty}\nn\\
&=&-{n\pi L_x\over 2}~\eta^2 \Big(f_n(r)\sqrt g~g^{rr}g^{xx}\partial_rf_n(r)\Big)_{r\rightarrow\infty}
\ee
Substituting the found solution $f_n(r)$ we obtain
\be
\log Z(n)&=&A_{\rm on-shell}=\pi  Lx ~\eta^2\left(-2 n (\log (n)+\log (r))+2 n \left(\psi^{(0)}\left(\frac{n}{2}\right)+\gamma \right)+2\right)+ {\cal O} \Big({1\over r}\Big)\nn\\
\ee
The terms linear in $n$ include divergent terms that should be subtracted [1]. However, they do not contribute to the entropy. 

The associated generalized gravitational entropy becomes 
\be
S_{GGE}&=&-n\partial_n\Big[\log Z(n)-n \log(1)\Big]_{n=1}={\pi  L_x \over 4}~\eta^2\Big(8-\pi^2\Big)
\ee
which is exactly the value in the real scalar field case.  The physical reason behind the coincidence is that in 3D the Maxwell field is dual to a real scalar field.  Therefore, the contribution of the electromagnetic field is just that from a  real scalar field in 3D.  

  Notice that only the asymptotical values of $\log Z(n)$ are same for real scalar field and Maxwell field.  The terms ${\cal O} \Big({1\over r}\Big)$ in each $\log Z(n)$ are different.  

\section {Backreaction of the Maxwell Field on the Area Law}
Now we will study the backreaction of the Maxwell field on the metric. The action is 
\be
-S=\int d^3x~\sqrt g~\Big[R-\Lambda -{1\over 4}F_{\mu\nu}F^{\mu\nu}\Big]
\ee
with $\Lambda=2$. The Einstein equations  are
\be
G_{\mu\nu}\equiv R_{\mu\nu}-{g_{\mu\nu}\over 2}\Big(R-2\Big)=T_{\mu\nu}
\ee
where 
\be
T_{\mu\nu}={1\over 4}g_{\mu\nu} F_{ab}F^{ab}-F_{\mu}^{~a}F_{a\nu}
\ee
Using the property shown in previous section that it remains only one component $A_x$ in Coulomb gauge we find that 
\be
T_x^x&=&{1\over 4} F_{ab}F^{ab}-F^{xa}F_{ax}\\
T_r^r&=&{1\over 4} F_{ab}F^{ab}-F^{ra}F_{ar}\nn=T_x^x-(\partial^\tau A^x)(\partial_tA_x)\nn\\
&=& T_x^x-{1\over r^2(1+r^2)}(\partial_\tau A_x)^2
\ee
To proceed we see that the backreaction of Maxwell field will  modify the  BTZ metric which can be written by [1]
\be
ds^2 ={dr^2\over 1+r^2+g(r)}+r^2d\tau^2+(1+r^2)(1+v(r))dx^2
\ee 
To the leading of perturbation (small $g(r)$ and $v(r)$)
\be
G_x^x&=&{ g'(r)\over  2r}\\
G_r^r&=&{g(r)\over 1+r^2}+{(1+r^2)v'(r)\over 2r}
\ee
Therefore, by using the Einstein equations we can find that 
\be
v'(r)=\Big({g(r)\over 1+r^2}\Big)'+{2\over r(1+r^2)^2}(\partial_tA_x)^2
\ee
Now we have a little trouble, because that the functions $g(r)$ and $v(r)$ are independent of $\tau$ while the time dependence of $A_x(\tau ,r)$ is $\cos(\tau)$, as used in previous section.  In fact, in this case we can first perform the integration over factor $\cos(\tau)$ within the function $(\partial_tA_x)^2$ in the action.  This integration gives  ${1\over 2\pi}\int_0^{2\pi}\cos(\tau)^2={1\over 2}$ in the action and, in conclusion, we can replace above $(\partial_\tau A_x)^2$ by ${1\over 2}A_x^2$.

Therefore, using the property that $g(0) = 0$ due to the regularity condition for the metric at the origin and $g(r)/r^2\rightarrow 0$ at infinity we find that 
\be
v(0)&=&\int_0^\infty dr {1\over r(1+r^2)^2}(A_x)^2
\ee
which implies that to the leading order of $\eta^2$ the deformation of area of horizon $\delta A$ is 
\be
S|_{\eta^2}&=&4\pi \delta A= -\eta^2~2\pi L_x  \int_0^\infty dr {1\over r(1+r^2)^2}(A_x)^2\approx -1.46838~\eta^2~L_x
\ee
after substituting the exact solution of $A_x$ (with $n=1$) found in previous section.

 We make two comments about our results.

1. The black hole formula is $S = 4\pi A = 4\pi A_0(1 + v(0))=4\pi (A_0 +\delta A)$ and $S|_{\eta^2}$ is the part of deformation by Maxwell field. 

2. The generalized gravitational entropy calculated in previous section is $S_{GGE}={\pi  L_x \over 4}~\eta^2\Big(8-\pi^2\Big)= -1.46838~\eta^2~L_x
$.  We thus see that $S_{GGE}=S|_{\eta^2}$. This shows that the backreaction of the Maxwell field on the area of horizon is equal to the associated  generalized gravitational entropy.

\section {Conclusion} 
After  Lewkowycz  and Maldacena [1] proposed the generalization of the usual black hole entropy formula [2-5]  several authors had studied the generalized gravitational entropy in higher-derivative gravity [8-14]. In this paper we focus on the fundamental property of  generalized gravitational entropy.  We first investigate the case of free real scalar field and then discuss the effect of the interacting scalar field. Next, we investigate the Maxwell field system. We are able to exactly solve the wave equation and calculate the analytic value of  the generalized gravitational entropy in Coulomb gauge.  We also use the Einstein equation to find the effect of backreaction of the Maxwell field on the geometry.  The associated modified area law is consistent with the generalized gravitational entropy.  We have shown a possible way to calculate the generalized gravitational entropy of the interacting scalar field and Maxwell field system in  Coulomb gauge.  We also explicitly  show that  effect of backreaction of the Maxwell field on the area of horizon is the generalized gravitational entropy.

Note that in this paper we adopt the Coulomb gauge because the associated solution could be found easily.  However, it is important to see that whether the property  also be shown in the general gauge.  Also, we shall study the property in the more realistic geometry to confirm the property of  the generalized gravitational entropy.  These problems are under investigation.
\\
\begin{center} {\bf REFERENCES}\end{center}
\begin{enumerate}
\item  A. Lewkowycz and J. Maldacena, ``Generalized gravitational entropy," JHEP 08 (2013) 090  [arXiv:1304.4926 [hep-th]].
\item   J. D. Bekenstein, ``Black Holes and Entropy", Phys. Rev. D 7, 2333 (1973).
\item J. M. Bardeen, B. Carter and S. W. Hawking, ``The four laws of black hole mechanics, "Commun. Math. Phys. 31, 161 (1973).
\item S.W. Hawking, ``Particle creation by black holesCommun," Math. Phys. 43, 199 (1975), [Erratum-ibid. 46, 206 (1976)]..
\item  G. W. Gibbons and S. W. Hawking, ``Action Integrals and Partition Functions in Quantum Gravit," Phys. Rev. D 15, 2752 (1977).
\item Jr., C. G. Callan and F. Wilczek, ``On geometric entropy", Phys. Lett., B333 (1944) 55, [arXiv:hep-th/9401072].
\item  S. N. Solodukhin, `` Entanglement entropy of black holes," Living Rev. Relativity 14, (2011), 8 090  [arXiv:1104.3712 [hep-th]].
\item A. Bhattacharyya, A. Kaviraj and A. Sinha, `` Entanglement entropy in higher derivative holography,"  JHEP 1308 (2013) 012 [arXiv:arXiv:1305.6694 [hep-th]].
\item  B. Chen and J-J Zhang, ``Note on generalized gravitational entropy in Lovelock gravity,"  JHEP 07 (2013) 185 [arXiv:arXiv:1305.6767 [hep-th]].
\item D. V. Fursaev, A. Patrushev and S. N. Solodukhin, ``Distributional Geometry of Squashed Cones,"  Phys. Rev. D 88 (2013)044054[arXiv:arXiv:1306.4000 [hep-th]].
\item A. Bhattacharyya, M. Sharma and A. Sinha, ``On generalized gravitational entropy, squashed cones and holography,"  JHEP 1401 (2014) 021 [arXiv:arXiv: 1308.5748 [hep-th]].
\item X. Dong, `` Holographic Entanglement Entropy for General Higher Derivative Gravity," JHEP 1401 (2014) 044 [arXiv:arXiv:1310.5713 [hep-th]].
\item J. Camps, ``Generalized entropy and higher derivative Gravity," JHEP 1403(2014) 070 [arXiv:arXiv:1310.6659 [hep-th]].
\item A. Bhattacharyya and M. Sharma, ``On entanglement entropy functionals in higher derivative gravity theories," [arXiv:arXiv:1405.3511 [hep-th]].
\end{enumerate}
\end{document}